\documentclass[%
 reprint,
 superscriptaddress,
 amsmath,amssymb,
 aps,
]{revtex4-1}

\usepackage{graphicx}
\usepackage{dcolumn}
\usepackage{bm}
\usepackage{bm}
\usepackage{amsmath,amsfonts,amssymb}
\usepackage{epstopdf}
\usepackage{xcolor}
\usepackage{color}
\usepackage{wasysym}
\usepackage{mathrsfs}

\begin{document}


\title{Deflection of a reflected intense circularly polarized light beam\\ induced by asymmetric radiation pressure}


\author{Y. H. Tang}
\affiliation{State Key Laboratory of Nuclear Physics and Technology, Peking University, Beijing 100871, China}

\author{Z. Gong}
\affiliation{State Key Laboratory of Nuclear Physics and Technology, Peking University, Beijing 100871, China}
\author{J. Q. Yu}
\affiliation{State Key Laboratory of Nuclear Physics and Technology, Peking University, Beijing 100871, China}
\author{Y. R. Shou}
\affiliation{State Key Laboratory of Nuclear Physics and Technology, Peking University, Beijing 100871, China}
\author{X. Q. Yan}
 \email[]{x.yan@pku.edu.cn}
 \affiliation{State Key Laboratory of Nuclear Physics and Technology, Peking University, Beijing 100871, China}
 \affiliation{CICEO, Shanxi University, Taiyuan, Shanxi 030006, China}
 \affiliation{Shenzhen Research Institute of Peking University, Shenzhen 518055, China}


\date{\today}

\begin{abstract}
A novel deflection effect of an intense laser beam with spin angular momentum is revealed theoretically by an analytical modeling using radiation pressure and momentum balance of laser plasma interaction in the relativistic regime, as a deviation from the law of reflection. The reflected beam deflects out of the plane of incidence with a deflection angle up to several milliradians, when a non-linear polarized laser, with the intensity $I_0\sim10^{19}$W/cm$^2$ and duration around tens of femtoseconds, is obliquely incident and reflected by an overdense plasma target. This effect originates from the asymmetric radiation pressure caused by spin angular momentum of the laser photons. The dependence of the deflection angle of a Gaussian-type laser on the parameters of laser pulse and plasma foil is theoretically derived, which is also confirmed by three dimensional particle-in-cell simulations of circularly polarized laser beams with the different intensity and pulse duration. 
\end{abstract}


\maketitle



The earliest record about reflection of light can be retrospected in around 200 B.C., when Greek physicist Archimedes defended his homeland Syracuse, utilizing the giant mirror to set fire to invader's ships\cite{heath2002works}. During the past two thousands years, humans have been persisting in pursuing the essential nature of reflection of light, which is summarized as the principle of interface conditions for electromagnetic fields in the classical optics\cite{jackson1999classical}. The reflection of light at a plane interface is one of the most basic optical processes, which is ubiquitous in general optical system and experimental facilities\cite{strickland1985compression,thaury2007plasma,balabanski2017new,chekhlov200635,miller2004national}. The fundamental case, reflection of a plane wave at a perfectly flat interface between two homogeneous isotropic media, is characterized by the law of reflection and Fresnel equations\cite{born1999principles}. The law of reflection claims that the direction of reflected light is in the plane of incidence, and the angle of reflection equals the angle of incidence. Nevertheless, the situation will becomes complicated when a realistic light beam possesses a finite spatial size and spectrum width, where the law of reflection is not enough for description. Neglecting shape deformations of the reflected beam, one can distinguish four basic deviations from the geometrical-optics picture. These deviations are usually referred as the spatial and angular Goos--H{\"a}nchen (GH)\cite{goos1947neuer,artmann1948berechnung,mcguirk1977angular,ra1973reflection,antar1974gaussian,chan1985angular} and Imbert--Fedorov (IF)\cite{fedorov1955,schilling1965strahlversetzung,imbert1972calculation} shifts following commonly recognized terminologies\cite{bliokh2013goos}.

\begin{figure}
\includegraphics[keepaspectratio=true,width=86mm]{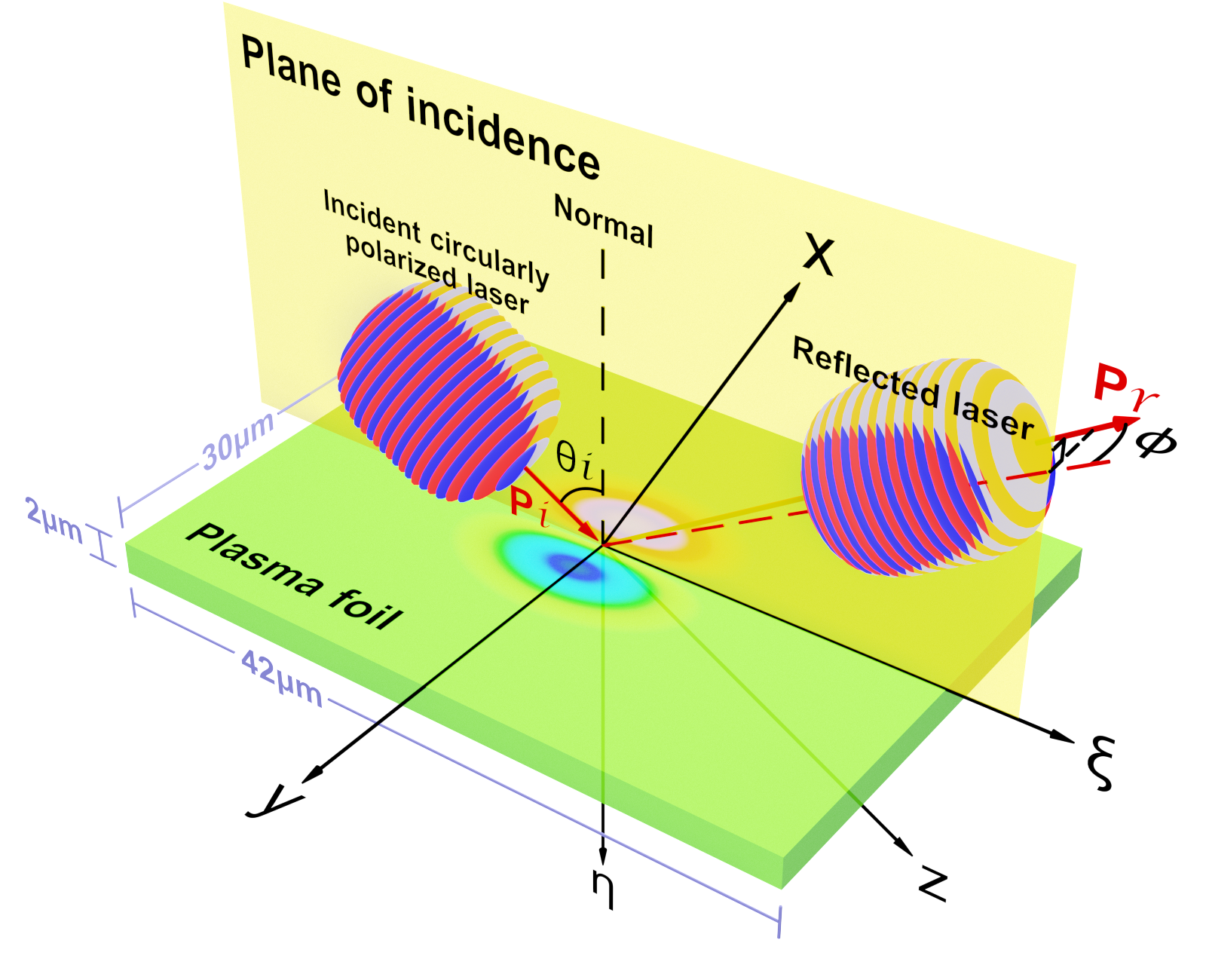}%
\caption{Schematic diagram of reflection of an intense circularly polarized laser beam at a plane surface $(\eta=0)$ of an overdense plasma foil. The coordinate frame $(x, y, z)$ is attached to the incident beam, while the frame $(\xi, y, \eta)$ is attached to the surface of the plasma foil. The laser pulse, propagating alone $z$-axis and being focused on the surface, is obliquely incident at an angle $\theta$ and reflected by the plasma foil. The reflected light deflects out of the plane of incidence $(y=0)$ at a deflection angle $\phi$. \label{fig:1}}
\end{figure}

The spatial GH shifts, originating from the dispersion of the reflection coefficients, displace the reflected beam within the plane of incidence\cite{artmann1948berechnung}. In contrast, the effect of spatial IF shifts, pertinent to the conservation of angular momentum of light beam\cite{player1987angular,fedoseyev1988conservation}, causes a displacement perpendicular to the plane of incidence to the reflected circularly polarized beam, which is also known as the spin Hall effect\cite{onoda2004hall}. All above basic shifts can occur in a generic beam reflection and have a consequent displacement at sub-wavelength scales. Apart from the spatial shifts, the angular shifts can be understood as shifts in the linear momentum space. The angular IF shift shows that the linear momentum of the reflected beam has a component perpendicular to the plane of incidence implying that the reflected beam deflects out of the plane of incidence, which is unveiled latterly\cite{bliokh2007polarization,bliokh2008geometrodynamics,hosten2008observation,aiello2008role,aiello2009duality,qin2011observation,bliokh2009goos,merano2010orbital}. 

However, these previous studies mainly focus on the weak light and the homogeneous isotropic media. With the progress of state-of-the-art laser technologies\cite{mourou2006optics}, sub Peta-Watt laser facilities with an intensity of $10^{20} \mathrm{W/cm^2}$ becomes available in practical experiments. When such an high-intensity pulse fires on the medium, it will ionize the materials into plasma within a temporal scale of femtoseconds, where the interaction between strong pulses and plasma is entirely different from that of the weak light. In the latest studies, a new deflection of the reflected intense vortex laser beam similar to the angular IF shift was proposed in Ref\cite{zhang2016deflection}, where the required vortex laser beam with relativistic intensity could be achieved via the theoretical proposed nonuniform reflection schemes\cite{shi2014light,gong2019angular} or other more robust experimental methods\cite{brabetz2015laser,denoeud2017interaction}.

In this Letter, leveraging on theoretical analyses and fully self-consistent three dimensional (3D) particle-in-cell (PIC) simulations, we present a novel phenomenon where the reflected beam deflects out of the plane of incidence when an intense non-linearly polarized laser beam without orbital angular momentum is obliquely incident and reflected by an overdense plasma foil. 
This effect is intrinsically induced by the antisymmetric radiation pressure originating from the spin angular momentum possessed by a non-linearly polarized laser beam.
Considering the balance between the radiation pressure and particles momentum flux, the dependence of the deflection angle $\phi$ on the laser pulse intensity $I_0$ and temporal duration $\tau$ is analytically derived and confirmed by the numerical results of 3D simulations. In addition, we discuss the influence of a preplasma with an exponential decayed density profile induced by a relatively low contrast of the incident laser pulse on this abnormal deflection phenomenon based on 3D PIC simulations with different plasma scale lengths.

\begin{figure}
\includegraphics[keepaspectratio=true,width=86mm]{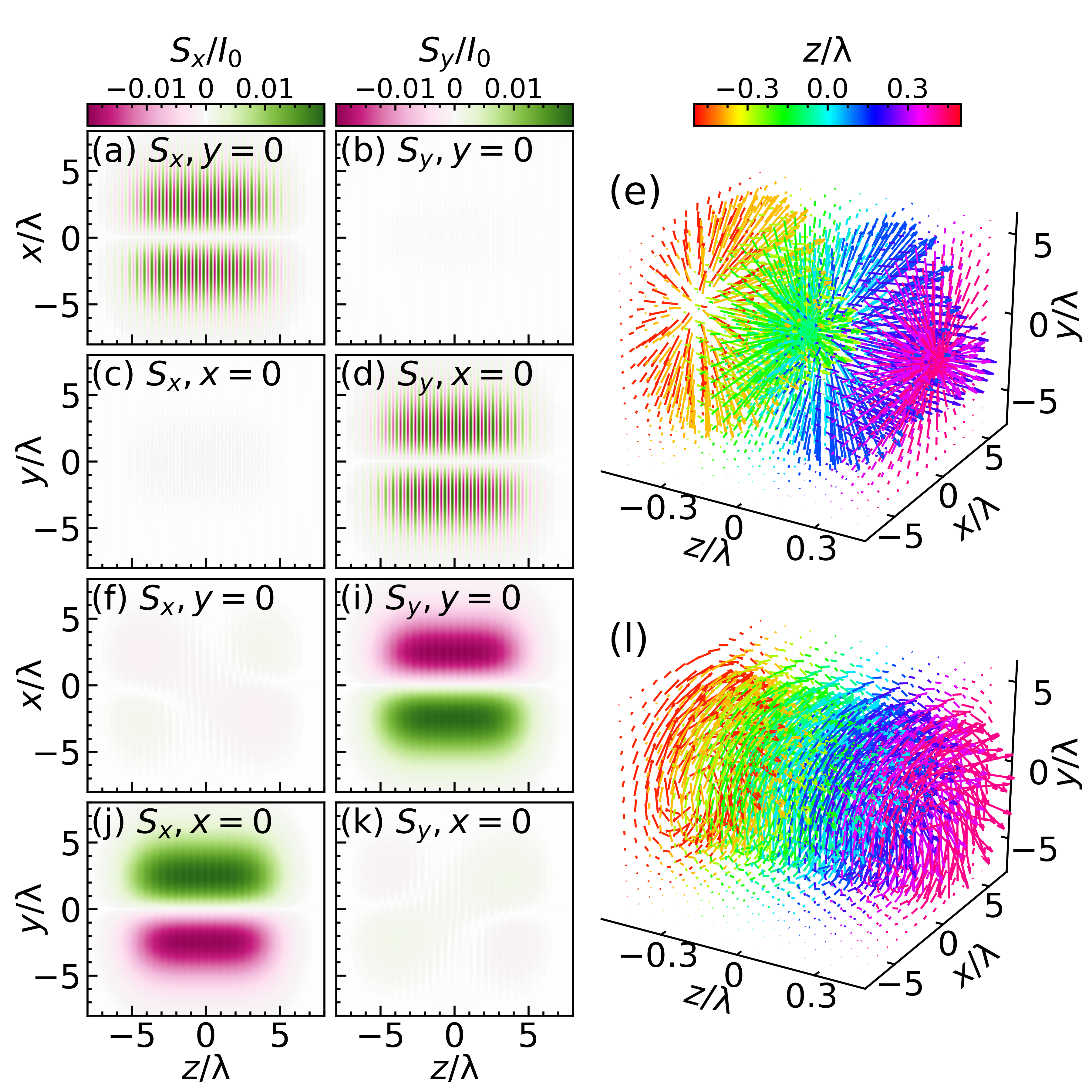}%
\caption{The Poynting vectors $\mathbf{S}$ of the tightly focused Gaussian laser beams in the numerical simulation. (a)-(e) correspond to the linear polarization, while (f)-(l) correspond to the circular polarization. (a) and (f) $S_x$ on the plane $y=0$, (b) and (i) $S_y$ on the plane $y=0$, (c) and (j) $S_x$ on the plane $x=0$, and (d) and (k) $S_y$ on the plane $x=0$. (e) and (l) are the schematic diagrams of the distribution of the Poynting vector, where the longitudinal component $S_z$ is two orders of magnitude scaled down to show the transverse component more clearly. $I_0$ is the peak intensity (i.e. $S_z$ at the focus of the laser beam) and $\lambda$ is the wavelength of the laser. \label{fig:new}}
\end{figure}

In order to investigate the radiation pressure exerted on foil surface by the incident light beam, we turn to the interaction process between intense laser beams and plasma, which is characterized by the synergy of Lorentz force and Maxwell's equations. Neglecting the trivial collisions in plasma, the dominant force density is written as $\mathbf{f} =\nabla \cdot \boldsymbol{\sigma} - \left (1/c^2 \right ) \partial \mathbf {S}/\partial t$, where $\mathbf{S}= \left (1/\mu_0 \right ) \mathbf  {E}\times \mathbf  {B}$ is the Poynting vector and $\boldsymbol {\sigma }$ is the Maxwell stress tensor. The components of $\boldsymbol {\sigma }$ read $\sigma_{ij}=\varepsilon_0 \left ( E_i E_j - \delta_{ij} E^2/2 \right ) + \left (1/\mu_0 \right ) \left ( B_i B_j -\delta_{ij} B^2/2 \right )$, where $\varepsilon_0$ the vacuum permittivity, $\mu_0$ the vacuum permeability and $\delta_{ij}$ Kronecker's delta. Integrating the force density $\mathbf{f}$ over the interacting region, the total Lorentz force is expressed as $\mathbf{F} = \oint_{a}\boldsymbol{\sigma}\cdot \mathrm{d}\mathbf{a} - (1/c^2) \mathrm{d}( \int_{V}\mathbf {S}\cdot \mathrm{d} V )/\mathrm{d}t $, where $\mathbf{a}$ is the area of boundary surface and $V$ is the closed volume, and Gauss's flux theorem is utilized when deriving the first term in $\mathbf{F}$. In practical interaction, as the laser pulse is totally reflected by the plasma target, only one side of foil's surface is irradiated by the electromagnetic pressure. The fields on other sides of surface and inside the foil are negligible, i.e., $\mathbf{S}\approx 0 $ and $\boldsymbol{\sigma}_\mathrm{inside}\approx0$. Consequently, we get the radiation pressure caused by the electromagnetic field on the foil, $\mathcal{P} = - \mathrm{d} \left( \mathbf{F} \cdot \hat{\mathbf{n}}\right )/\mathrm{d} s = - \left ( \boldsymbol{\sigma}\cdot \hat{\mathbf{n}}  \right ) \cdot \hat{\mathbf{n}}$, where $\hat{\mathbf{n}}$ is the unit normal vector of the foil surface. In the coordinate frame $(x,y,z)$ of incident laser beam shown in Fig.~\ref{fig:1}, $\hat{\mathbf{n}} = \left (\sin \theta_i, 0, -\cos \theta_i \right )$, the pressure can be written as
\begin{equation}
\mathcal{P}_{las} = - \left (\sigma_{xx} \sin^2\theta_i + \sigma_{zz} \cos^2\theta_i - \sigma_{xz}\sin2\theta_i \right ).
\label{eq:pressure_laser}
\end{equation}

The concrete form of $\mathcal{P}_{las}$ can be determined if the Maxwell stress tensor $\sigma_{ij}$ of laser beam is explicitly described. In the realm of intense laser plasma interaction, a beam with a Gaussian envelop is commonly recognized as an approximate form of laser beam. Under the paraxial approximation, an eigen solution of the Maxwell equation in vacuum space prescribes the complex vector potential as $\mathbf{A}= A_0 \left (w_0/w \right ) \exp \left [- \left (x^2+y^2 \right )/w^2 \right ] \exp \{\mathrm{i} [\omega t-kz - k \left (x^2+y^2 \right )/2R + \psi(z) ] \} \left (e_x \hat{\mathbf{x}} + e_y \hat{\mathbf{y}} \right )$, where $A_0$ the amplitude constant, $w_0$ the waist radius, $w = w_0\sqrt{1 + \left (z/z_R \right )^2}$ and $R(z) = z \left [ 1 + \left (z_R/z \right )^2 \right ]$. $z_R = \pi w_0^2/\lambda$ the Rayleigh length and $\lambda$ the laser wavelength. $k = \omega/c$ the wave number, $\hat{\mathbf{x}}$ and $\hat{\mathbf{y}}$ unit vectors, $e_x$ and $e_y$ the relative amplitude satisfying $ \left |e_x \right |^2 + \left |e_y \right |^2 = 1$ and $\psi(z) = \arctan \left( z/z_R \right)$ the Gouy phase. The equation indicates that the wave propagates along the $z$-axis and the beam is focused at $(0,0,0)$. Leveraging on the Lorenz gauge condition, the electromagnetic fields can be derived from $\mathbf{E} = -\mathrm{i}\omega \left [\mathbf{A}+ \left (c^2/\omega^2 \right )\nabla\nabla\cdot\mathbf{A} \right ]$ and $\mathbf{B} = \nabla\times\mathbf{A}$. After substituting the $\mathbf{E}$ and $\mathbf{B}$ into the formula of $\boldsymbol{\sigma}$ and assuming $\nabla\nabla\cdot\mathbf{A}\ll k^2\mathbf{A}$, we obtain the components of the effective Maxwell stress tensor
\begin{equation}
\begin{aligned}
\left \langle {\sigma}_{xx} \right \rangle &= - \frac{1}{4} \varepsilon_0 A_0^2 \omega_0^2 \frac{ x^2 + y^2 }{ z^2 + z_R^2 } \exp\left (-2\frac{x^2+y^2}{w_0^2}\right), \\
\left \langle {\sigma}_{zz} \right \rangle &= - \frac{1}{4} \varepsilon_0 A_0^2 \omega_0^2 \left ( 2 - \frac{ x^2 + y^2 }{ z^2 + z_R^2 } \right ) \exp\left (-2\frac{x^2+y^2}{w_0^2}\right), \\
\left \langle {\sigma}_{xz} \right \rangle &= - \frac{1}{2} \varepsilon_0
A_0^2 \omega_0^2 \frac{ zx + {s} z_R y }{ z^2 + z_R^2 } \exp\left (-2\frac{x^2+y^2}{w_0^2}\right) .
\label{eq:MST_solution}
\end{aligned}
\end{equation}
Here $\left\langle\ \right\rangle$ denotes the temporal average over one laser period and ${s} = 2 \Im \left (e_x^* e_y \right )$ is the beam's helicity which represents its intrinsic spin angular momentum pertinent with the polarization states. ${s} = 0$ corresponds to the linear polarization while ${s} = 1\ (-1)$ corresponds to the right (left) handed circular polarization. Substituting Eq.~(\ref{eq:MST_solution}) into Eq.~(\ref{eq:pressure_laser}), we obtain the averaged radiation pressure of the incident beam

\begin{eqnarray}
\left \langle {\mathcal{P}_{las}} \right \rangle &=& \frac{2I_0}{c} e^{ -{2 \left (x^2+y^2 \right )}/{w_0^2} } \left (  \cos^2 \theta_i  - s \frac{\lambda y }{ \pi w_0^2 } \sin 2\theta_i \right ),
\label{eq:pressure_mean}
\end{eqnarray}
where non-grazing incident is assumed and high order terms are neglected. A factor of 2 is multiplied in Eq.\eqref{eq:pressure_mean} due to the effect of the reflected beam and $I_0 = \varepsilon_0 c \omega^2 A_0^2/2$ is the peak intensity of the laser beam. The first term in Eq.\eqref{eq:pressure_mean} exhibits the isotropic properties which is symmetric along the transverse ($y$-axis) direction. However, the second term, induced by spin polarization, is proportional to the coordinate $y$ which definitely results in an asymmetric distribution of radiation pressure with respect to the central axis $y=0$.

The asymmetric radiation pressure can be physically explained by the spin angular momentum possessed by the non-linearly polarized laser beam. 
To demonstrate it more clearly, we numerically simulate the tightly focused Gaussian laser beams by self-consistent PIC simulations, and show the comparison of the Poynting vector of linear and circular polarization in Fig.~\ref{fig:new}. The Poynting vector $\mathbf{S}$ represents the momentum density $\mathbf{g}$ since $\mathbf{g} = \mathbf{S}/c^2$. For linear polarization, the transverse component is radial and oscillating. For circular polarization, the transverse component is circumferential and continuous, and that's the origin of the spin angular momentum. In other words, a non-linearly polarized beam with a finite spatial size possesses a screwed direction distribution of the momentum density. The circumferential component of momentum density increases with the decreasing of the beam radius according to the conservation of angular momentum. When such a tightly focused beam obliquely impinges the plasma, the angle between the momentum density and the normal direction at a point is slightly different to the corresponding point with respect to the plane of incidence, which indicates the asymmetry of the normal component of the momentum density i.e. radiation pressure. 
The asymmetry is particularly apparent for a small waist radius of the laser beam, and is then transferred to the plasma by the complex coupling mechanisms between laser and plasma, further transferred to the reflected beam by the motion of plasma\cite{chopineau2019identification}. Consequently, the symmetry is neither persisted in laser field nor in plasma. The asymmetric radiation pressure shown in Eq.~(\ref{eq:pressure_mean}) causes asymmetric equilibrium position of the plasma surface, and it's equivalent to tilt the surface to one side of the plane of incidence, so the reflected beam is predicted to deflect out of the plane of incidence in this case.

\begin{figure}
\includegraphics[keepaspectratio=true,width=86mm]{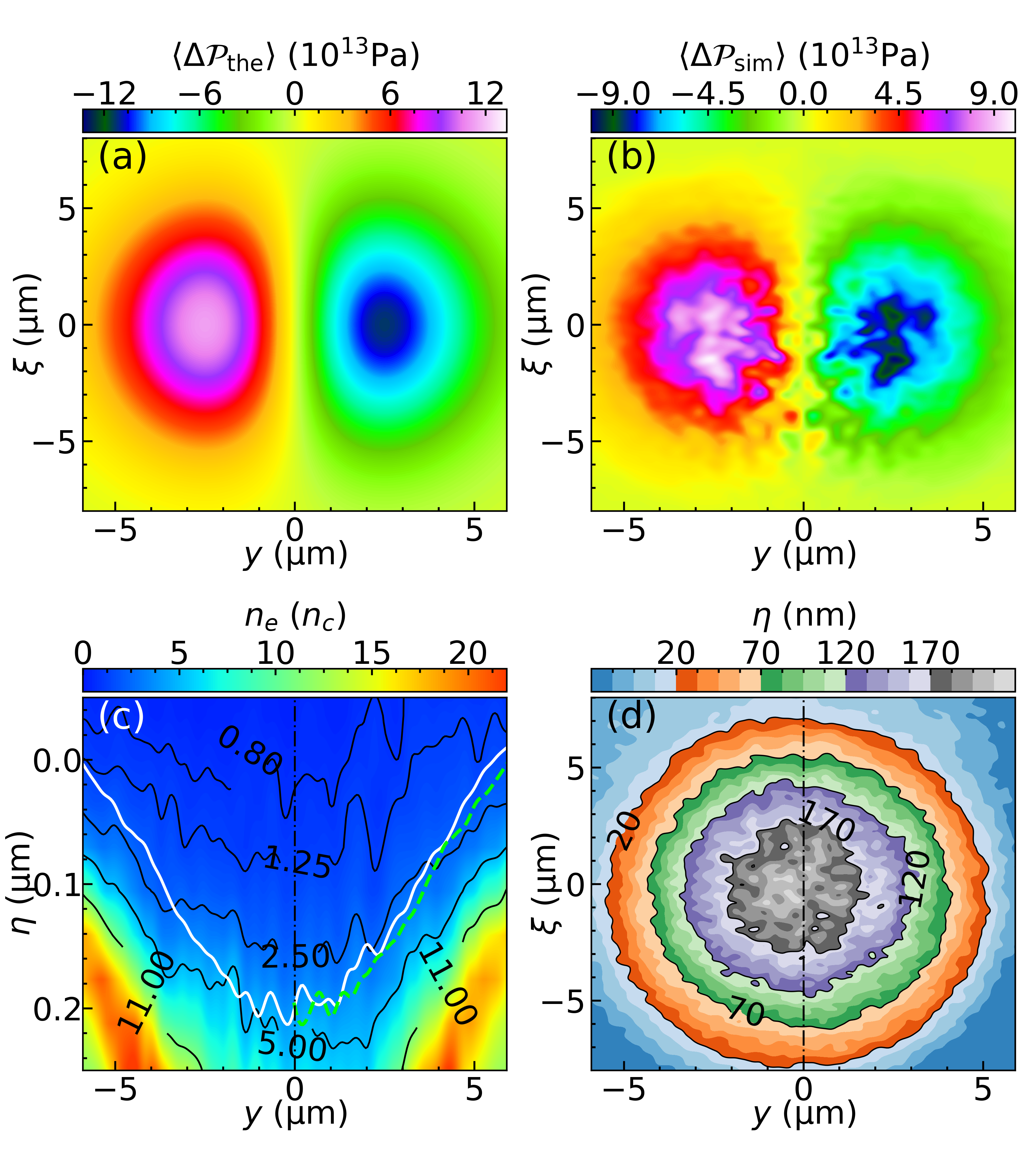}%
\caption{(a) The temporally averaged radiation pressure difference between $(\xi,y,0)$ and $(\xi,-y,0)$ in theoretical model. (b) The temporally averaged pressure difference over one laser period at $t=100\mathrm{fs}$ in simulation, when the half of laser duration finishes being reflected by the surface. (c) The electron density in simulation at $\xi = 0$ and $t=150\mathrm{fs}$. The black lines are the isodensity lines, while the white solid line is the relativistic surface and the green dashed line is the axisymmetric part of white solid line with respect to $y=0$ for easy comparison. (d) The contour of the relativistic surface in simulation at $t=150\mathrm{fs}$ when the reflection is finished.\label{fig:2}}
\end{figure}

To prove our theoretical analyses, self-consistent 3D PIC simulations are performed via code EPOCH\cite{arber2015contemporary}. The simulation region is a box with size of $50 \mathrm{\mu m} \times 40 \mathrm{\mu m} \times 40 \mathrm{\mu m}$, which is uniformly divided into $1000 \times 800 \times 800$ cells in $x \times y \times z$ direction. A circularly polarized (${s}=1$) Gaussian beam with a temporal profile of $e^{-(t-t_0)^4/\tau_0^4}$ is focused on the surface of foil, as shown in Fig.~\ref{fig:1}. The beam possesses a wavelength $\lambda = 1.0 \mathrm{\mu m}$, a waist radius $w_0 = 5.0 \mathrm{\mu m}$, duration $\tau_0 = 20 \mathrm{fs}$ and a peak intensity $I_0 = 4.38 \times 10^{19} \mathrm{W/cm^2}$. The intensity corresponds to a normalized field amplitude $a_0 = e A_0/m_e c = 5.66$ (for circularly polarized beam, the actual amplitude is $a_0/\sqrt{2} = 4.0$ ). For simplicity, the incident angle is set as $\theta_i=\pi/4$ and the beam propagates along the $z$-axis, so that the reflected beam should propagate along the $x$-axis according to the law of reflection. The foil is a hydrogen plasma with an electron density of $12n_c$ where $n_c = \varepsilon_0 m_e\omega^2/e^2 = 1.1\times10^{21} \mathrm{cm^{-3}}$ is the critical density\cite{gibbon2005short}. The foil with a size of $42 \mathrm{\mu m} \times 30 \mathrm{\mu m} \times 2 \mathrm{\mu m}$ is put in $\xi \times y \times \eta$ direction (as shown in Fig.~\ref{fig:1}) and the particles are represented by 16 macro-particles per cell for both electron and proton. The center of the irradiated surface is set coincidentally with the origin of the coordinate frame.

Induced by the antisymmetric radiation pressure with respect to the plane of incidence $y=0$ (the second term in Eq.~\eqref{eq:pressure_mean}), the symmetry of the plasma surface is deteriorated. The temporally averaged radiation pressure difference exerted on the initial plasma surface $\eta=0$, $\left<\Delta \mathcal{P}(\xi,y)\right>$, is shown in Fig.~\ref{fig:2}, where the theoretical derived $\left<\Delta  \mathcal{P}_{\mathrm{the}}\right>=- \left (2I_0s\lambda y/\pi w_0^2c \right ) \exp[-2(\xi^2\cos^2\theta_i+y^2)/w_0^2] \sin2\theta_i$ is drawn in (a) while the pressure difference of 3D PIC simulation $\left<\Delta \mathcal{P}_{\mathrm{sim}}\right>$ is visualized in (b). The result in Fig.~\ref{fig:2}(b) corresponds temporally averaged value over one laser period at time $t=100\mathrm{fs}$ in simulation, when the half of laser duration finishes being reflected by the surface. 
The simulation result is qualitatively in good agreement with the theoretical result, but has a little difference on scale due to the theoretical simplification of the radiation pressure of the reflected beam.

When the high-intensity pulse irradiates on the plasma surface, the critical density for laser pulse being able to propagate through will be extended by the electrons' effective Lorentz factor $\gamma_\mathrm{eff}$, which is also termed as relativistic induced transparency\cite{kaw1970relativistic,lefebvre1995transparency,shen2001transparency,willingale2009characterization}. The asymmetric electron density distribution caused by the asymmetric pressure, is exhibited in Fig.~\ref{fig:2}(c) where the white solid line denotes the contour of relativistically correct critical density $n_e = \gamma_\mathrm{eff}n_c$ and $\gamma_\mathrm{eff}=\sqrt{1+a^2}=\sqrt{1+16\exp{\left [-2 \left (\xi^2\cos^2\theta_i+y^2 \right )/w_0^2 \right ]}}$. The green dashed line at $y>0$ is the axisymmetric part of white solid line at $y<0$ with respective to the plane of incidence $y=0$, which demonstrates a pronounced depth difference. The specific depth $\eta$ of the critical density surface $n_e=\gamma_\mathrm{eff}n_c$ is plotted in Fig.~\ref{fig:2}(d), which indicates the left part is more deep than the right part as a result of the asymmetric radiation pressure.

\begin{figure}
\includegraphics[keepaspectratio=true,width=86mm]{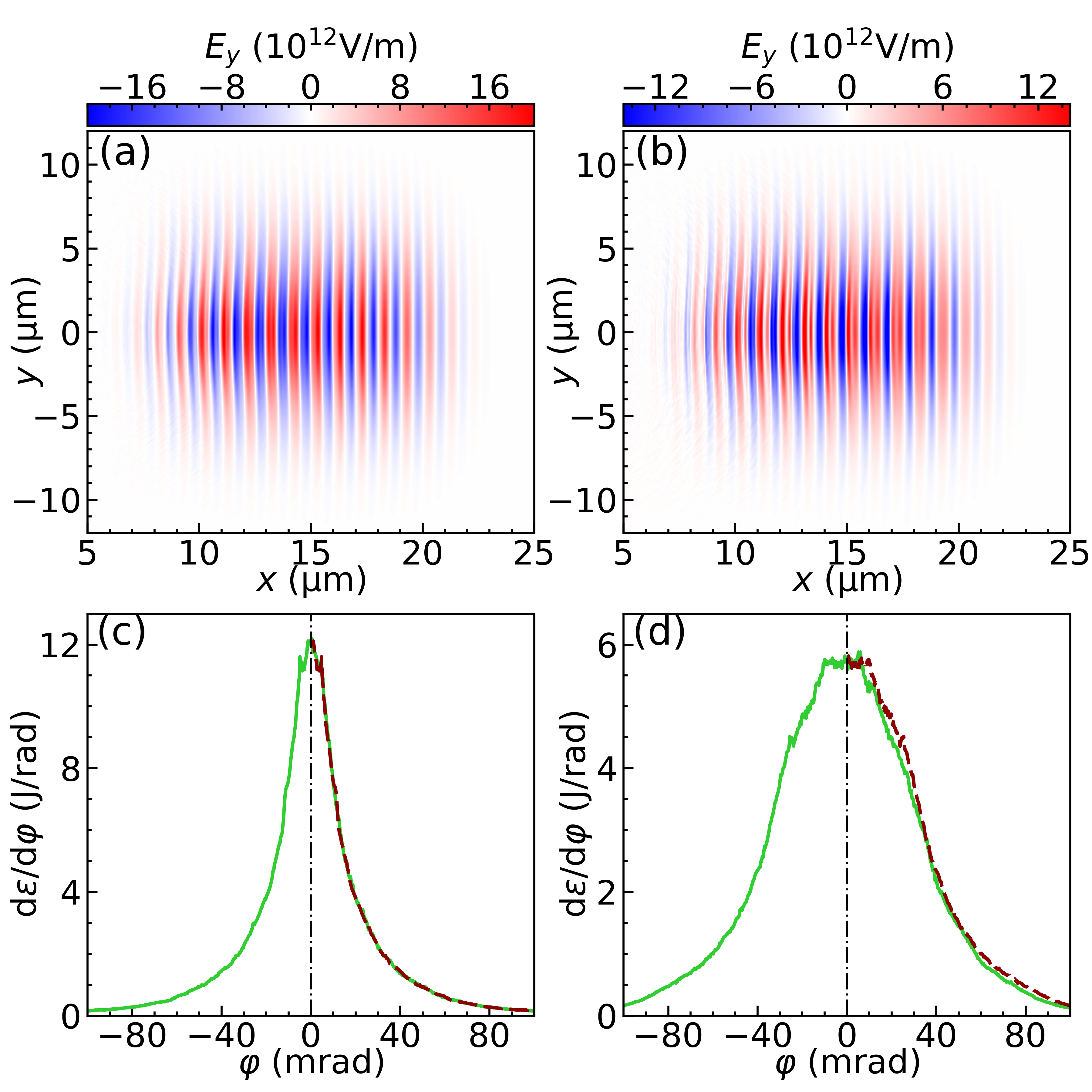}%
\caption{ The electric field and angular profile of the reflected laser beam at $t=150\mathrm{fs}$. (a) and (c) are the result of linear polarization, while (b) and (d) are the circular polarization. $\varepsilon$ is the total laser energy and $\varphi$ is the angle between the laser momentum and the plane of incidence. The red dashed lines ($\varphi>0$) in (c) and (d) are the axisymmetric part of the green solid lines ($\varphi<0$) with respect to $\varphi = 0$. \label{fig:3}}
\end{figure}

Due to this asymmetric radiation pressure, the reflected beam deflects out of the plane of incidence with an angle of mrad magnitude, as shown in Fig.~\ref{fig:3} where the comparison to the case of a linear polarization is supplemented. It is hard to directly distinguish the deflection angle of mrad magnitude from the electromagnetic field in Fig.~\ref{fig:3}(a) and (b), but there is an alternative indirect method to demonstrate the deflection of the circular polarization. The local angle $\varphi$ between the laser momentum and the plane of incidence can be obtained from the equation $\varphi = g_y/\left | \mathbf{g} \right | = S_y/\left | \mathbf{S} \right |$. The laser energy is $\varepsilon = \int \left ( \varepsilon_0 E^2/2 + B^2/2 \mu_0  \right )\mathrm{d}V$. Then the angular profiles Fig.~\ref{fig:3}(c) and (d) are obtained by making histogram in the whole simulated space for different $\varphi$ with the laser energy as its weights, and it's easy to observe the difference between the linear and circular polarization. 
For the linear polarization, the whole coincidence, occurring between the axisymmetric part with respect to $\varphi = 0$ of $\mathrm{d}\varepsilon/\mathrm{d}\varphi$ (the red dashed line) and the original part (the green  solidline), shows that the reflected beam exactly propagates parallel to the plane of incidence. In contrast, for the circular polarization, the displacement of mrad magnitude between the axisymmetric and original part indicates that the entire beam has a angular shift along the direction of negative $y$-axis.

\begin{figure}
\includegraphics[keepaspectratio=true,width=86mm]{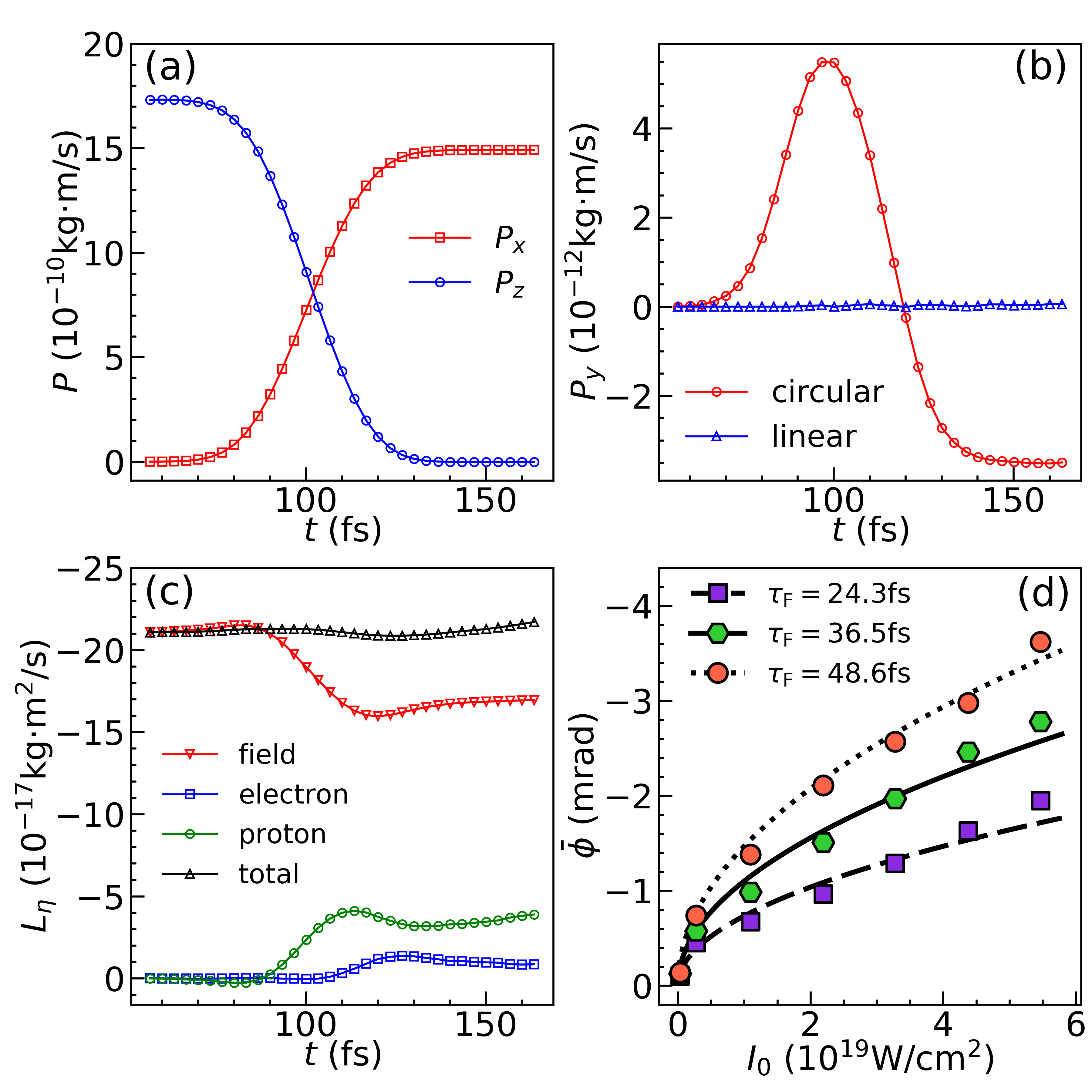}
\caption{(a)-(c) The temporal evolution of linear momentum $\mathbf{P}$ and angular momentum $\mathbf{L}$ for the entire space in simulation. (a) $P_x$ (red squared line), $P_z$ (blue circular line) of the field for the circular polarization. (b) $P_y$ of the field for p-linear polarization (blue triangular line) and right circular polarization (red circular line). (c) $L_\eta$ the normal component of angular momentum for the circular polarization. (d) Comparison of the deflection angles of the simulation results (the points) and the theoretical model (the lines) with different laser intensity and pulse duration.\label{fig:4}}
\end{figure}

Quantitatively, the mean deflection angle can be derived from the $\bar{\phi} = P_y/\left |\mathbf{P} \right |$ where $P$ represents the total linear momentum of the reflected beam\cite{bliokh2013goos}.
So the linear momentum of the electromagnetic field $\mathbf{P} = \varepsilon_0 \int_V \left ( \mathbf{E} \times \mathbf{B} \right ) \mathrm{d}V$ is numerically integrated over the entire 3D simulation domain and the result is shown in Fig.~\ref{fig:4}(a), where $P_z$ represents the momentum of the incident light while $P_x$ implies that of the reflected beam. Starting from $t \simeq 60\mathrm{fs}$, $P_z$ quasi linearly drops down to zero until the $t \simeq 140\mathrm{fs}$, indicating that the laser beam is totally reflected by the surface.
The difference between final $P_x$ and initial $P_z$ implies that nearly 14\% laser momentum is transferred into plasma particles.
In Fig.~\ref{fig:4}(b), the transverse momentum $P_y$ of the circularly polarized beam, exhibits the nonconservative feature during the reflection, which directly proves the projection of laser propagating direction is no longer zero in the transverse direction $y$-axis, and the angle can be estimated as $\bar{\phi} \simeq P_y/\left | P_x \right | = -2.46\mathrm{mrad}$ when the reflection terminates and the momentum of the entire space represents that of the reflected beam at $t=150\mathrm{fs}$.
In realistic experiments, the detection of an angular change as small as $\mathrm{\mu rad}$ magnitude has been achieved\cite{gray2001laser,genoud2011active,iwasinska2014system}, so it's reasonable for us to claim that the observation of this abnormal deflection effect in real experiments is achievable. The $P_y$ of the p-linearly polarized beam with the same intensity is also shown in Fig.~\ref{fig:4}(b), and it's two orders of magnitude smaller than the circularly polarized beam.

To exclude the influence of IF effect, which is also able to result in a transverse deflection in surface reflection, we leverage on the angular momentum of the electromagnetic wave $\mathbf{L} = \varepsilon_0 \int_V \mathbf{r} \times \left ( \mathbf{E} \times \mathbf{B} \right ) \mathrm{d}V$. The temporal evolution of the normal component of angular momentum is illustrated in Fig.~\ref{fig:4}(c). The angular momentum transferred to electrons and protons implies the broken of rotational symmetry with respect to the normal direction, so that the angular momentum of electromagnetic field is no longer conserved which manifests intrinsic difference with respect to IF effect\cite{fedoseyev2009conservation,bliokh2009goos}.

In order to determine which parameters impact on the deviating angle of the reflected light, an analytical modeling based on the momentum balance is investigated. The relation between the mean deflection angle $\bar{\phi}$ and the final oblique angle $\alpha$ of the asymmetric deformation of plasma surface is estimated as $\bar{\phi} \simeq \alpha$. 
It is worth pointing out that the surface of the plasma synchronously oscillates with the laser field, and the maximum displacement of the oscillation of surface is far larger than the final deformation (e.g. the deformation shown in Fig.~\ref{fig:2}(c) and (d)), when an intense laser beam is obliquely incident and reflected by a plasma. Therefore, the temporally averaged deformation is considered to derive the final deformation of the surface. The momentum balance of $\eta$-axis in the instantaneous rest frame, written as $\left \langle \mathcal{P}_{las} \right \rangle = 2 n_i m_i \left \langle v_{\eta} \right \rangle^2$, is taken into account, where $n_i$ is the ion density, $m_i$ is the ion mass and $v_\eta$ is the velocity of the plasma surface in the lab frame\cite{robinson2009relativistically}.
When the beam is not a grazing incidence (i.e. $\left |  2 \lambda s y \tan \theta/\pi w_0^2 \right | \ll 1$), the velocity can be written as
\begin{equation}
\left \langle v_{\eta} \right \rangle = \sqrt{\frac{I_0}{n_i m_i c}}e^{-(x^2+y^2)/w_0^2} \left ( 1 - \frac{\lambda s y}{\pi w_0^2} \tan \theta \right ).
\label{eq:velocity}
\end{equation} 
We can find that for a non-linearly polarized beam, the plasma surface has different temporally averaged velocity with respect to the plane of incidence. The velocity of one side is larger than the other side, and it causes the surface tilted to one side. After the laser interaction with a duration $\tau$, there is a displacement difference $\Delta \eta= \left \langle \Delta v_\eta \right \rangle \tau$ between surface points $(x_s,\pm y_s)$ where $\Delta v_\eta$ is the velocity difference. The oblique angle of the plasma surface can therefore be expressed as $\alpha = \Delta \eta/ 2y_s $. Using Eq.~(\ref{eq:velocity}), we can obtain the oblique angle as
\begin{equation}
\alpha = - \frac{\lambda \tau s \tan \theta}{\pi w_0^2} \sqrt{\frac{I_0}{n_i m_i c}}e^{-(x_s^2+y_s^2)/w_0^2}.
\label{eq:oblique_angle}
\end{equation}
Via accounting for the spatial average effect, the deflection angle of the reflected beam can be derived as
\begin{equation}
\bar{\phi} \simeq - \frac{\lambda \tau s \tan \theta}{\pi w_0^2} \sqrt{\frac{I_0}{n_i m_i c}}.
\label{eq:deflection_angle}
\end{equation}

A series of 3D PIC simulations with different laser intensity and pulse duration are performed to further testify Eq.~(\ref{eq:deflection_angle}). The numerical deflection angles in each simulations, calculated via $\bar{\phi} = P_y/\left | P_x \right |$, are shown in Fig.~\ref{fig:4}(d) as solid dots. The theoretical lines are obtained from Eq.~(\ref{eq:deflection_angle}) by replacing pulse duration $\tau$ with the full width half maximum duration $\tau_\mathrm{F}=2\sqrt[4]{\ln{2}}\tau_0$ in simulation. As shown in Fig.~\ref{fig:4}(d), all lines are close to the simulation data points after multiplied by a same scale coefficient $0.62$. It demonstrates the relation $\bar{\phi} \propto \sqrt{I_0}$ and $\bar{\phi} \propto \tau$, which indicates that the effect is only pronounced for an intense laser and the deflection angle is increased with the raise of laser pulse duration.

\begin{figure}
\includegraphics[keepaspectratio=true,width=68mm]{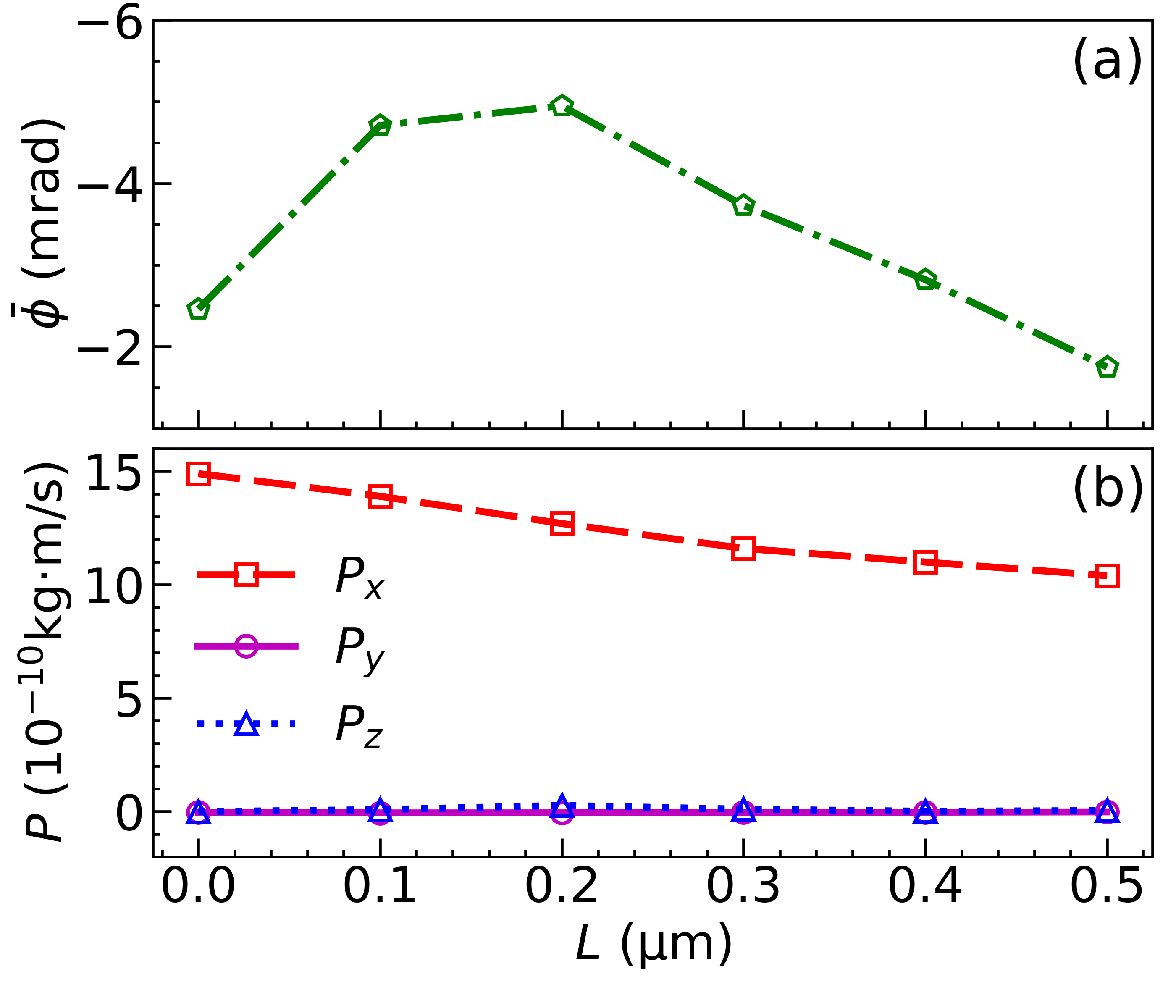}%
\caption{ The dependence of deflection angle (a) and the linear momentum (b) of the reflected laser beam on different plasma scale length $L$ at simulation time $t=150\mathrm{fs}$. $L=0$ is the case of plasma target with a cut-off density profile.\label{fig:length}}
\end{figure}

Considering the impact of the prepulse of a realistic laser pulse in the experiment, the effect is investigated on exponentially decayed density profile of different characteristic scale length $L$ in simulations. Specifically, the distribution of the plasma density can be described in the form $n_e= {n_0}/(1+e^{-\eta/L})$ where $n_0=12n_c$ corresponds to the density of the cut-off boundary, and other parameters are the same as the initial simulation. The scale length is a simplification of the actual conditions observed in experiments and is commonly utilized when considering pre-plasmas induced by the prepulse\cite{peebles2017investigation}. 
The results of deflection angle and linear momentum of the reflected laser beam are shown in Fig.~\ref{fig:length}, where $|P_z| \ll |P_x|$ indicates that the laser is still totally reflected and the reflected laser beam nearly propagates alone $x$-axis. With the increase of the scale length, the $|\mathbf{P}|$ decreases appreciably due to the energy dissipation of laser in the plasma, while the magnitude of the deflection angle increases first and then decreases. For a relatively long scale length, the laser is reflected before it reaches the maximum density area, so it is equivalent to decrease the $n_i$ in Eq.~(\ref{eq:deflection_angle}), which can explain the first increase of the magnitude of the deflection angle. And that also gives the simulations the significance to the experiment, although the maximum plasma density in experiments is usually several hundred $n_c$, much larger than $12n_c$ used in the simulations. However, Eq.~(\ref{eq:deflection_angle}) is no longer valid for a too long scale length since the boundary can't be simply treated as an interface any more. 

In conclusion, a novel deflection effect of an intense laser with spin angular momentum is revealed by theoretical model and simulation, as a deviation from the law of reflection. The reflected beam deflects out of the plane of incidence with an experimentally observable deflection angle, when an intense non-linear polarized laser is obliquely incident and reflected by an overdense plasma target. The analytical modeling is set up in the relativistic regime of laser plasma interaction, shows the asymmetric radiation pressure of a laser beam possessing spin angular momentum, and indicates the rotational symmetry breaking of the foil and the deflection of the reflected laser beam by momentum balance.
A formula is given to predict the deflection angle of a Gaussian-type laser and reveal the relation between the angle and the parameters of laser and plasma. Finally, a succession of full 3D PIC simulations of circularly polarized laser beams with different laser intensity and pulse duration demonstrate the relation between the angle and the laser intensity as well as the pulse duration, 
and another succession of full 3D PIC simulations of different plasma scale length demonstrate the existence of the deflection when the incident laser pulse possesses a low contrast.

\begin{acknowledgments}
The PIC code EPOCH was in part funded by the UK EPSRC grants EP/G054950/1, EP/G056803/1, EP/G055165/1 and EP/M022463/1.
Simulations were supported by High-performance Computing Platform of Peking University.
\end{acknowledgments}

\bibliographystyle{apsrev4-1}

\end{document}